\documentclass[pre,twocolumn,nofootinbib,superscriptaddress,amsmath, amssymb,aps,balancelastpage,longbibliography]{revtex4-2}
\usepackage{graphics}
\usepackage{graphicx}
\usepackage{picture}
\usepackage{placeins}
\usepackage{float}
\DeclareGraphicsExtensions{.pdf,.eps} 
\usepackage{amsmath,scalefnt}
\usepackage{mathtools}
\usepackage{amssymb}
\usepackage{verbatim}
\usepackage{amsmath,amsfonts,bbm}
\usepackage{amsbsy}
\usepackage{color}
\usepackage{cancel}
\usepackage{soul}
\usepackage{dsfont}
\usepackage[svgnames]{xcolor} 
\usepackage[breaklinks=true,colorlinks=true,linkcolor=DarkBlue,urlcolor=DarkBlue,citecolor=DarkBlue]{hyperref}
\usepackage{subfigure}
\usepackage{braket}
\usepackage{longtable}     
\usepackage{calrsfs}
\usepackage{eucal}
\usepackage{bbold}
\usepackage{multirow}
\usepackage{url}           
\usepackage{bm}            
\usepackage{microtype}
\usepackage{array}
\usepackage{makecell}
\usepackage{enumitem}
\usepackage{hhline}
\usepackage{tabularx}
\usepackage{xfrac}

\DeclareSymbolFont{matha}{OML}{txmi}{m}{it}
\DeclareMathSymbol{\varv}{\mathord}{matha}{118}

\def\ba#1{\left(\begin{array}{#1}}
\def\ea{\end{array}\right)}
\def\bsm{\left(\begin{smallmatrix}}
\def\esm{\end{smallmatrix}\right)}

\renewcommand{\rm}{\mathrm}

\usepackage[normalem]{ulem}

\begin{document}

\title{Probing lambda-gravity with Bose-Einstein condensates}

\author{Hector A. Fernandez-Melendez}
\email{H.A.Fernandez-Melendez@soton.ac.uk}
\affiliation{School of Physics and Astronomy, University of Southampton, Southampton SO17 1BJ, United Kingdom}
\author{Alexander Belyaev}
\email{A.Belyaev@soton.ac.uk}
\affiliation{School of Physics and Astronomy, University of Southampton, Southampton SO17 1BJ, United Kingdom}
\affiliation{Particle Physics Department, Rutherford Appleton Laboratory, Chilton, Didcot, Oxon OX11 0QX, UK}

\author{Vahe Gurzadyan}
\email{gurzadyan@yerphi.am}
\affiliation{Center for Cosmology and Astrophysics, Alikhanian National Laboratory and Yerevan State University, Yerevan 0036, Armenia}

\author{Ivette Fuentes}
\email{I.Fuentes-Guridi@soton.ac.uk}
\thanks{Previously known as Fuentes-Guridi and Fuentes-Schuller.}
\affiliation{School of Physics and Astronomy, University of Southampton, Southampton SO17 1BJ, United Kingdom}
\affiliation{Keble College, University of Oxford, Oxford OX1 3PG, United Kingdom}

\date{\today}

\begin{abstract}
We propose a precise test of two fundamental gravitational constants using a detector concept that exploits the dynamics of quantum phononic excitations in a trapped Bose-Einstein condensate (BEC), operable at the scale of tabletop experiments. In this setup, the sensitivity is enhanced by approximately 2 orders of magnitude through the use of a tritter operation, which mixes phononic excitations with the BEC's ground state. The BEC exhibits unique sensitivity to the two key components of the gravitational potential in $\Lambda$-gravity: the Newtonian $GM/r$ term and the cosmological constant $\Lambda r^2$, both entering the most general function following from a Gurzadyan's theorem. Using state-of-the-art experimental design, we predict that the gravitational constant $G$ could be measured with an accuracy up to $10^{-17}$ N m$^2$/kg$^2$, representing an improvement by 2 orders of magnitude over current measurements. Moreover, this experiment aims to establish the best Earth-based upper limit on $\Lambda$ at $<10^{-31}$ m$^{-2}$, marking the first laboratory-based probe of the cosmological constant. Additionally, the setup allows for the measurement of the distance-dependent behavior of each term in the gravitational potential, providing a means to test modified gravity theories.
\end{abstract}

\maketitle


\section{Introduction}

The current standard model of cosmology, the $\Lambda$CDM model, aligns remarkably well with main cosmological data \cite{Peebles2025}, yet the nature of both dark matter and dark energy remains unexplained.

A key component of the $\Lambda$CDM model -- dark energy -- is often associated with the cosmological constant, $\Lambda$. Experimentally, the existence of the $\Lambda$ term is well established within the standard $\Lambda$CDM framework.

Gurzadyan's theorem \cite{G1} shows that the most general force law consistent with the equivalence of gravitational forces from a spherical mass and a point mass includes a $\Lambda$ term in the weak-field limit of General Relativity:
\begin{equation}
 F = - \frac{GMm}{r^2} + \frac{\Lambda r m c^2}{3} \ .
 \label{eq:lambda}
\end{equation}  
This form fits local Universe observations \cite{G2,GS1}, describes galaxy groups and clusters dynamics \cite{GS2,GS4}, and suggests resolving the Hubble tension via local and global flows \cite{R1,R2,R3,GS4,GS5,GS6}.

Note that, Eq.~\ref{eq:lambda} satisfies the first condition of Newton's shell theorem (sphere-point equivalency) but not the second one (force-free inside a spherical shell). The second term in Eq.~\ref{eq:lambda} implies a non-zero force inside a shell, consistent, e.g., with spiral galaxy disk observations shaped by spherical halos \cite{Kr}. Thus, according to this theorem, gravity is described by two constants, $G$ and $\Lambda$, the latter recognised by Einstein as a universal constant \cite{E1,E2}. Its consequences were studied in \cite{GS3}, and it plays a key role in Conformal Cyclic Cosmology, enabling rescaling of physical constants between aeons \cite{P,GP1,GP2,GS3}. 

It is therefore a principal issue if this theorem can also be tested or at least constrained in a laboratory experiment. We propose tabletop experiments with trapped Bose-Einstein condensates (BECs) to probe both fundamental constants of $\Lambda$-gravity and the distance dependence in Eq.~\ref{eq:lambda}. Despite quantum sensors -- ultraprecise sensitivities \cite{Belenchia2022Quant} -- tests of modified gravity in such experiments are limited. In contrast, quantum proposals to search for dark matter -- via optomechanics \cite{Carney2021Mecha}, atom interferometry \cite{Graham2016DarkM}, and BECs \cite{Howl2023Quant} -- are more common.

Theoretical studies indicate that collective excitations in trapped BECs are highly sensitive to gravitational effects \cite{Sabin2014Phono, ahmadi_relativistic_2014, sabin_thermal_2016, Ratzel2018Dynam, Howl2023Quant, hartley_quantum-enhanced_2024}. At nanokelvin temperatures, BEC excitations behave as free quasiparticles (phonons) \cite{Dalfovo1999Theor, Pitaevskii2003BoseE}, responsive to gravitational field changes. Quantum metrology techniques, with entangled phononic states, can achieve sensitivities surpassing solid-state systems \cite{bose2023massivequantumsystemsinterfaces}. Resonant effects to periodic gravitational changes can further enhance sensitivity \cite{Ratzel2018Dynam, Sabin2014Phono, sabin_thermal_2016, Bravo2020Gravi}.

A recent theoretical proposal uses phonon states in a frequency interferometric protocol enhanced by a tritter operation, mixing phonon modes with the BEC ground state to improve sensitivity scaling as $1/\sqrt{N_p N_0}$, where $N_p$ is the number of squeezed phonons and $N_0$ the number of ground-state atoms \cite{Howl2023Quant,Szigeti2017Pumpe}. This outperforms earlier gravitational-wave detection proposals \cite{Sabin2014Phono}.

We propose a frequency interferometric method to measure $G$ and $\Lambda$ using a small oscillating mass near the BEC. The tritter enhances sensitivity by about 2 orders of magnitude beyond prior Newtonian acceleration measurements \cite{Ratzel2018Dynam}. Phonon squeezing of $\sim30.4$ dB could yield gravitational acceleration sensitivities as low as $10^{-18}$ m/s$^2$, enabling measurement of $G$ with relative accuracy $\sim10^{-7}$, 2 orders better than current best values. This experiment could also set the first Earth-based upper bound on $\Lambda < 10^{-31}\ \mathrm{m}^{-2}$. Given that BEC spin squeezing routinely reaches 6--8 dB in laboratories \cite{SU11BEC1,Xin2023LongL,Oberthaler2015Twist}, this precision appears achievable \cite{Juschitz2021Twomo,Gross2012SpinS,Tufarelli2012Input}.

The next section describes the experimental setup. The third section introduces the quantum metrology framework and the predicted BEC sensitivity to gravity. The fourth section presents the estimated experimental precision. The final section concludes.

\section{Gravity from an Oscillating Mass}
\label{sec:exp-setup}

The gravitational force on a mass $m$ is defined by the potential $\phi$ as
\begin{align}
 \Vec{F}_{G} = - m \Vec\nabla \phi,
 \label{eq:force}
\end{align}
where $\phi$ is given by Eq.~\ref{eq:lambda}:
\begin{align}
 \phi = -\frac{MG}{r} - \frac{\Lambda r^{2} c^{2}}{6}.
 \label{eq:h00}
\end{align}

We consider an experiment with an oscillating sphere of mass $M \sim 100$ g and frequency $\Omega$ placed near a BEC of length $L$, confined in a uniform box trap \cite{Gaunt2013BECof}, aligned with the oscillation direction. The BEC (size $L \simeq 100\,\mu\text{m}$) is positioned at distance $R_0 \gg L$ from the source mass, which oscillates with amplitude $\delta_R \simeq 1$ mm. The source mass radius is about 1--2 cm, consistent with a tungsten sphere of radius $\sim 1.1$ cm and mass near 100 g, realistic for laboratory conditions.
\begin{figure*}[tb] 
 \center
 \includegraphics[width=0.7\textwidth]{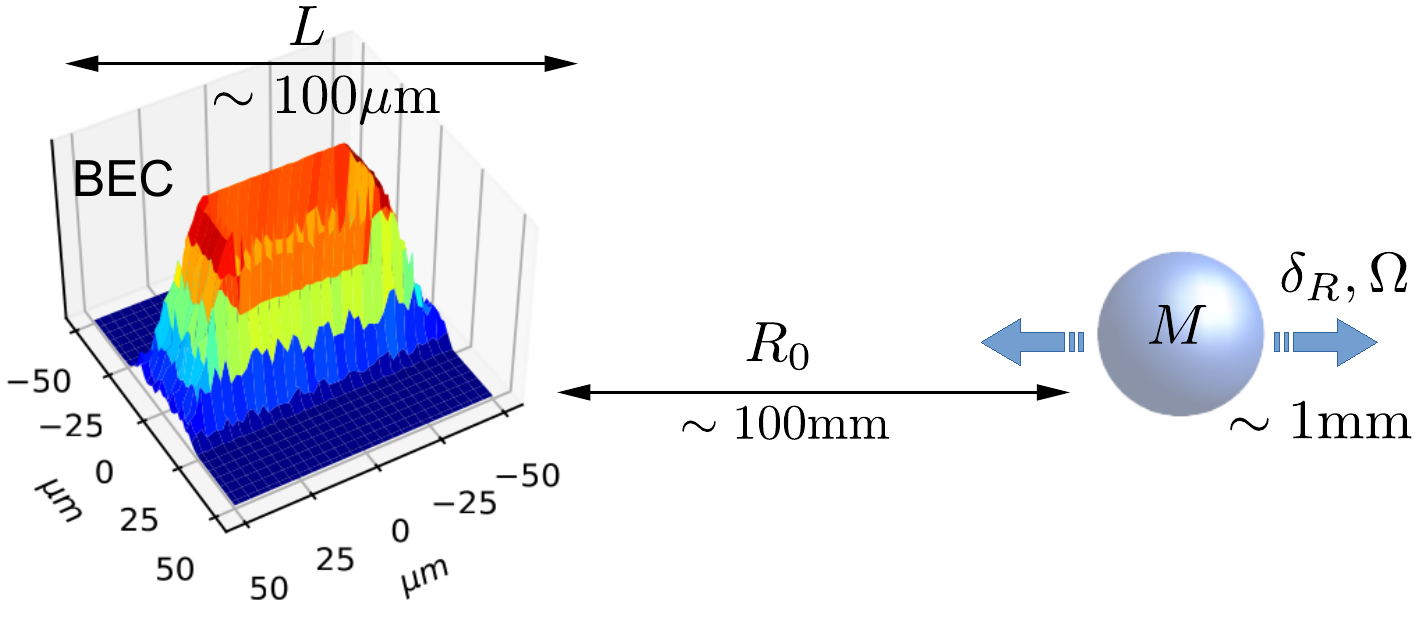}
 \caption{Experimental setup: an oscillating sphere of mass $M$, frequency $\Omega$, and amplitude $\delta_R$ (right) at distance $R_0 \gg \delta_R$ from a BEC of length $L$ (left).}
 \label{fig:bec}
\end{figure*}
The distance between the oscillating mass center and a point in the BEC at coordinate $x \in [0,L]$ is
\begin{equation}
 r(x,t) = R_0 + \delta_R \sin(\Omega t) + x \equiv R_0 (1 + \Delta_{x,t}),
 \label{eq:rxt}
\end{equation}
where $\Delta_{x,t} = [\delta_R \sin(\Omega t) + x]/R_0$ with $|\Delta_{x,t}| \ll 1$ for expansion.

As derived in Appendix~\ref{supp:grav-pot}, the amplitude of the time-dependent acceleration on the BEC, $a^\text{BEC}$ -- our key observable -- is approximately
\begin{equation}
 a^\text{BEC} \simeq \frac{2 \delta_R}{R_0^2} \left( \frac{MG}{R_0} + \frac{\Lambda R_0^2 c^2}{6} \right),
 \label{eq:abec}
\end{equation}
where $\mathcal{O}(\frac{\delta_R^2}{R_0^2})$ corrections to the Newtonian term are neglected for simplicity but can be included as needed. We use this simplified form for $a^\text{BEC}$ throughout the paper.

\section{Quantum metrology}
\label{sec:QuantMet}

Quantum metrology provides strategies to optimise precision in estimating a parameter $\epsilon$ encoded by a unitary transformation $\hat{U}(\epsilon)$ in a quantum state~\cite{Giovannetti2006Quant, Paris2009Quant, Giovannetti2011Advan, Ahmadi2014Relat}. Given the initial probe state, the optimal precision is set by the quantum Cramér-Rao bound (QCRB):
\begin{equation}
 \label{V.01}
 \Delta \hat{\epsilon} \geq \dfrac{1}{\sqrt{N_m \mathcal{F}(\epsilon)}} \, ,
\end{equation}
where $N_m$ is the number of measurements and $\mathcal{F}(\epsilon)$ is the quantum Fisher information (QFI)~\cite{Cramer1946Mathe, Braunstein1994Stati}. The QCRB bounds all positive-operator-valued measurements and can be saturated as $N_m \rightarrow \infty$. When the optimal measurement is unfeasible, suboptimal schemes like heterodyne detection may be used~\cite{Giovannetti2011Advan}.

The QFI quantifies distinguishability between states differing infinitesimally in $\epsilon$. For Gaussian states, it is conveniently computed via the covariance matrix formalism (CMF), where a Gaussian state is fully described by its displacement vector $\bm{d}$ and covariance matrix $\bm{\Gamma}$. In the complex representation~\cite{Safranek2015Quant, Arvind1995TheRe}, they read:
\begin{subequations}
 \begin{gather}
  \bm{d} \equiv \braket{\hat{\textbf{A}}} , \\
  \Gamma_{i j} \equiv \braket{\hat{\textbf{A}}_{i}\hat{\textbf{A}}_{j}^{\dagger} + \hat{\textbf{A}}_{j}^{\dagger}\hat{\textbf{A}}_{i}} - 2\braket{\hat{\textbf{A}}_{i}}\braket{\hat{\textbf{A}}_{j}^{\dagger}} \, ,
 \end{gather}
\end{subequations}
where $\braket{\ }$ denotes expectation with respect to $\hat{\rho}$ and $\hat{\textbf{A}} \equiv (\hat{A}_{1},\ldots,\hat{A}_{N};\hat{A}_{1}^{\dagger},\ldots,\hat{A}_{N}^{\dagger})^{T}$ is a $2N$-vector of bosonic annihilation/creation operators. Using CMF is natural here since Bogoliubov transformations (Appendix~\ref{supp:QFBEC}) are Gaussian and act as symplectic matrices $\bm{S}(\epsilon)$ on $\bm{d}$ and $\bm{\Gamma}$:
\[
\bm{d}' (\epsilon) = \bm{S}(\epsilon) \bm{d} (0), \quad \bm{\Gamma}' (\epsilon) = \bm{S}(\epsilon) \bm{\Gamma}(0) \bm{S}^{\dagger}(\epsilon) \,.
\]
In this formalism, the QFI takes the form~\cite{Monras2013Phase, Pinel2013Quant}:
\begin{align}
 \mathcal{F}(\epsilon) = \frac{1}{4}\text{Tr}\left[\left(\boldsymbol{\Gamma}(\epsilon)^{-1}
 \dot{\boldsymbol{\Gamma}}(\epsilon)\right)^2\right] + 2\dot{\boldsymbol{d}}^\dag(\epsilon) \boldsymbol{\Gamma}^{-1}(\epsilon) \dot{\boldsymbol{d}}(\epsilon),
\end{align}
where dots denote derivatives with respect to $\epsilon$.

\begin{figure}[ht] 
 \centering
 \includegraphics[width=0.48\textwidth]{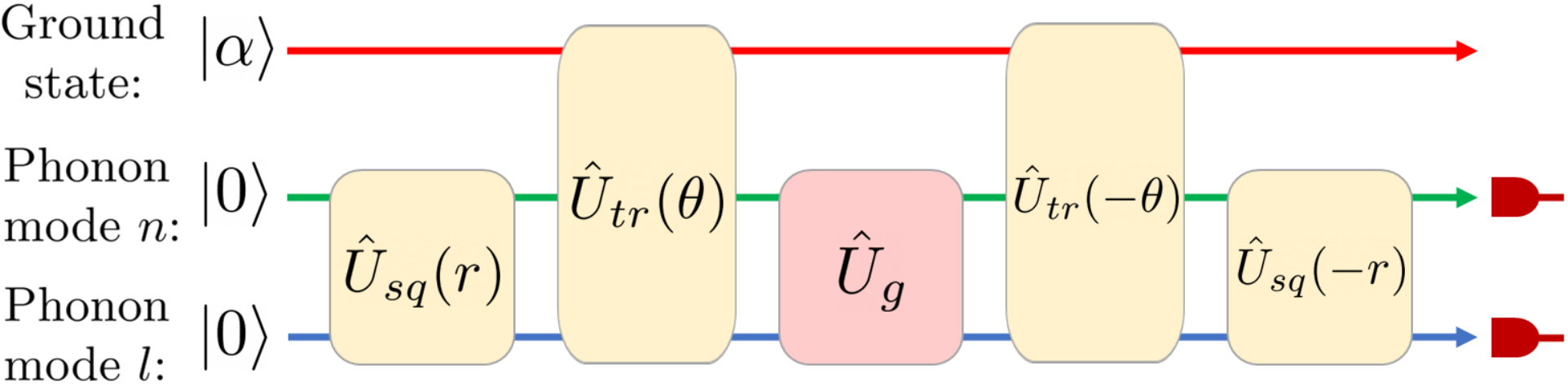}
 \caption{Probe state preparation. Two phonon modes (green and blue), initially in the vacuum state, undergo squeezing $\hat{U}_{\textrm{sq}}(r)$ and are then mixed with the BEC ground state (red) via the tritter $\hat{U}_{\textrm{tr}}(\theta)$. The gravitational parameter is encoded by $\hat{U}_{g}(a^\mathrm{BEC})$. The inverse transformations are then applied, followed by phonon number measurement.}
 \label{fig:freqint}
\end{figure}

We implement a three-mode frequency interferometry scheme~\cite{Howl2023Quant}, illustrated in Fig.~\ref{fig:freqint}, involving the BEC ground state and two phonon modes $n,l$ whose sum resonates with the mass oscillation frequency.

The BEC ground state approximates a coherent state $\hat{a}_{0} \ket{\alpha} = \alpha_{0} \ket{\alpha}$. Since squeezed states are optimal probes for Bogoliubov transformations~\cite{Safranek2016Optim, Monras2006Optim}, we apply two-mode squeezing on phonon modes:
\[
\hat{U}_{\textrm{sq}}(r) = e^{r(\hat{b}_{n}^{\dagger}\hat{b}_{l}^{\dagger} - \hat{b}_{n}\hat{b}_{l})},
\]
with squeezing parameter $r = |r| e^{i\vartheta_{\textrm{sq}} }$ and phase $\vartheta_{\textrm{sq}} \in \mathbb{R}$. This populates the modes parametrically. Sensitivity is enhanced by mixing the coherent state and squeezed phonons via the tritter transformation $\hat{U}_{\textrm{tr}}(\theta)$, generated by~\cite{Szigeti2017Pumpe}
\begin{equation}
 \hat{H}_{\textrm{tr}} (\theta) = \frac{\hbar \theta}{\sqrt{2}} \left[ e^{i \vartheta} \hat{a}_{0}^{\dagger} (\hat{b}_{n} + \hat{b}_{l}) + e^{- i \vartheta} \hat{a}_{0} (\hat{b}_{n}^{\dagger} + \hat{b}_{l}^{\dagger}) \right],
\end{equation}
with $\theta, \vartheta \in \mathbb{R}$. In the Bogoliubov approximation, $\hat{a}_{0}$ is replaced by $\sqrt{N_{0}}$. We assume the condensate remains largely undepleted by the squeezing and tritter transformations, with $N_0 = |\alpha_0|^2$. For this, $\theta$ must be sufficiently small~\cite{Howl2023Quant}.

When the oscillating mass is active, gravity acts on phonons via the two-mode squeezing $\hat{U}_{g}$ (Eq.~\ref{eq:UT} in Appendix~\ref{supp:QFBEC}), encoding the parameter $a^\mathrm{BEC}$. 
The interferometry closes with inverse transformations $\hat{U}_{\textrm{sq}}(-r)$ and $\hat{U}_{\textrm{tr}}(-\theta)$. Although less commonly discussed, inverse squeezing (antisqueezing) has been theoretically proposed and experimentally demonstrated in systems such as spinor BECs~\cite{Quant2016Linnemann, Quant2022Mao} and trapped ions~\cite{Exper2024Burd}. These show that inverse squeezing can be realized by reversing nonlinear evolution, approximately restoring the initial state. For phonons, this reversal may be achieved by inverting the modulation phase of the trapping potential. The inverse tritter transformation is analogous to time-reversing a beam splitter in linear optics; the details of its realisation can be found in Appendix~\ref{sec:inverse-tritter}.

Finally, releasing the trap allows single-atom detectors to measure atomic velocities, enabling estimation of the phonon number in the final state~\cite{Jaskula2012Acous}.

Taking optimal phases $\vartheta_{\textrm{sq}} = \pi/2$ and $\vartheta = \pi/4$, and assuming large $N, \bar{N}, r$, the QFI is
\begin{equation}
 \mathcal{F}(a^\mathrm{BEC}) \approx 8 (|M_{nl}|t/\hbar)^2 \theta^{2} N_0 N,
 \label{eq:bec-sens}
\end{equation}
quantifying the probe state's sensitivity to the gravitational force of the oscillating mass.

\section{Experimental details and results}
\label{sec:results}

To estimate the sensitivity to the minimal acceleration $\Delta a^{\rm BEC}$, we consider a 1D $^{87}$Rb Bose-Einstein condensate (BEC) trapped in a uniform potential, as described in \cite{Gaunt2013BECof}. The atomic mass is $m = 1.44 \times 10^{-25}$~kg, and the scattering length is $a_{sl} = 99\,r_B$ \cite{Egorov2013Measu}, where $r_B \simeq 5.29 \times 10^{-11}$~m is the Bohr radius. 

To ensure phonons propagate in one dimension, we impose the condition on the width-to-length ratio $\alpha_{\mathrm{WL}} \leq 0.1$ \cite{Shammass2012Phono}. Typical BEC lengths $L$ range between 50~$\mu$m and 1000~$\mu$m \cite{Kraemer2004Optim,vanderStam2007Large}, and the number of atoms $N_0$ varies from $1.6 \times 10^{3}$ to $1.1 \times 10^{9}$ \cite{Hansel2001BECon,Meppelink2010Therm,Fried1998BECof}.

The maximum experimental duration $t$ is limited by phonon and BEC half-lives, primarily determined by three-body recombination, with the time scale given by
\begin{equation}
 t_{\mathrm{hl}} = \frac{3}{2 D n_0^{2}},
\end{equation}
where $D=5.8 \times 10^{-30}$~cm$^6$s$^{-1}$ for $^{87}$Rb \cite{Burt1997Coher}. We select phonon modes with low quantum numbers to optimise sensitivity and ensure experimental feasibility \cite{Howl2023Quant}.

Following the analysis in \cite{Juschitz2021Twomo}, we set the squeezed phonon number to $N_p=1100$, corresponding to 30.4~dB of squeezing. Although the experimental generation of phonon squeezing remains challenging \cite{Jaskula2012Acous,Steinhauer2016Obser}, theoretical proposals suggest that such levels could be achievable under optimised conditions \cite{Gross2012SpinS,Tufarelli2012Input}. 

Several experimental conditions are maintained to validate our approximations:
\begin{itemize}
 \item Dilute gas regime: $n_0 |a_{sl}|^3 \ll 1$,
 \item Bogoliubov approximation: $N_{\mathrm{exc}} \ll N_0$, where $N_{\mathrm{exc}} \approx \frac{m c_s^{2}}{\hbar \omega_n} N_p$ is the number of excited atoms,
 \item Phonon regime: $\hbar \omega_{l,n} \ll m c_s^{2}$,
 \item Mode parity condition for resonance: $l + n$ is odd,
 \item Low temperature condition: $k_B T \ll \mu$, achievable down to 0.5~nK \cite{Steinhauer2014Obser},
 \item Tritter angle $\theta$ satisfies inequality (F3) in \cite{Howl2023Quant}.
\end{itemize}

The number of measurements performed is $N_m = \tau / t$, where $\tau$ is the total integration time. Combining Eq.~\eqref{eq:bec-sens} with Eq.~\eqref{V.01}, the sensitivity is
\begin{equation}
 \label{eq:delta-a}
 \Delta a^{\mathrm{BEC}} \approx \frac{\alpha_{\mathrm{WL}} \hbar \pi^{3} \sqrt{2 n l} (l^{2} - n^{2})^{2}}{16 m N_0 \theta \sqrt{L a_{sl} \tau t N_p} (l^{2} + n^{2})} \,.
\end{equation}

\begin{figure}[htbp] 
 \centering
 \includegraphics[width=0.48\textwidth]{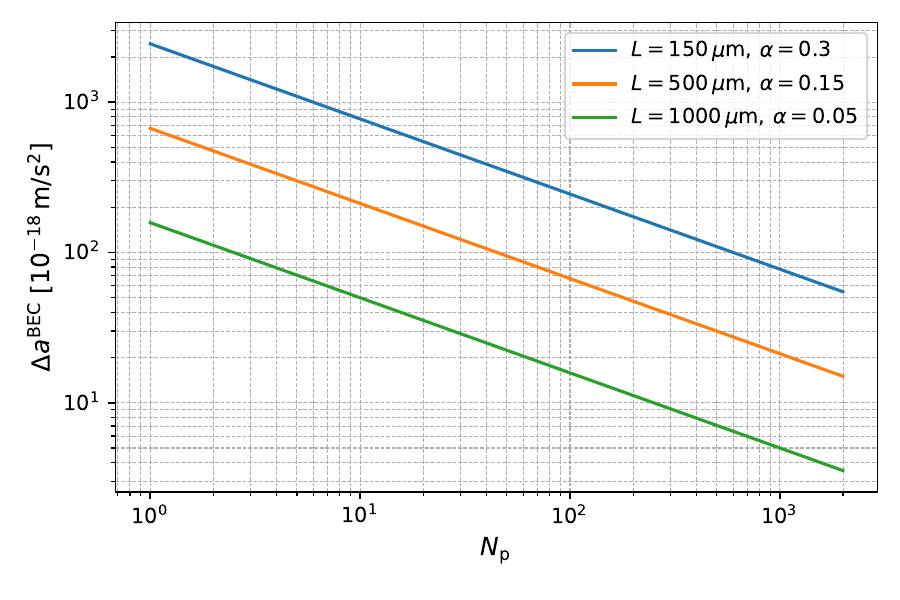}
 \caption{Sensitivity $\Delta a^{\mathrm{BEC}}$ as a function of the number of initial squeezed phonons $N_p$. The plot shows $\Delta a^{\mathrm{BEC}}$ calculated using Eq.~\eqref{eq:delta-a} for varying BEC lengths, with $N_p$ ranging from 1 to 1100. Increasing $N_p$ corresponds to enhanced sensitivity to $a^{\mathrm{BEC}}$.}
 \label{fig:delta-a-np}
\end{figure}

Table~\ref{table1} summarises the key parameters and resulting sensitivities for the $^{87}$Rb BEC setup. The speed of sound is $c_s = 1.9$~mm/s for a density $n_0 = 10^{14}$~cm$^{-3}$, resulting in a phonon frequency $\Omega = \pi c_s (n + l)/L = 17.7$~Hz for the chosen mode numbers. In addition, in Fig.~\ref{fig:delta-a-np} we illustrate $\Delta a^\text{BEC}$ dependence on $N_p$ in the case of a less optimistic number of phonons. One can see from this figure that sensitivity 
degrades by a factor of 30 if the number of phonons decreases by 3 orders of magnitude, e.g. from 1000 to 1. Recent work aiming at demonstrating entanglement produced a single phonon in a two-mode squeezed state using a BEC of 3000 helium atoms \cite{Obser2025Westbrook}. Theoretical studies indicate that achieving the sensitivities shown in Fig.~\ref{fig:delta-a-np} requires around 1000 squeezed phonons, which in turn demands a condensate of approximately $10^{9}$ atoms \cite{Juschitz2021Twomo,Gross2012SpinS,Tufarelli2012Input}. Current experimental efforts are focused on reaching this regime.

\begin{table}[htbp]
 \centering
 \begin{tabular}{lll}
  \hline
  Parameter & Symbol & Value/range \\
  \hline
  BEC length ($\mu$m) & $L$ & 150, 500, 1000 \\
  Width-to-length ratio & $\alpha_{\mathrm{WL}}$ & 0.3, 0.15, 0.05 \\
  Number of atoms & $N_0$ & $10^{9}$ \\
  Squeezed phonon number & $N_p$ & 1100 \\
  Phonon mode numbers & $l,n$ & 1, 2 \\
  Single-run duration (s) & $t$ & 1 \\
  Integration time (days) & $\tau$ & 60 \\
  Atomic mass (kg) & $m$ & $1.44 \times 10^{-25}$ \\
  Scattering length (m) & $a_{sl} = 99 r_B$ & $5.24 \times 10^{-9}$ \\
  Tritter angle (rad) & $\theta$ & 0.31 \\
  \hline
  Sensitivity $\times 10^{-18}$ & $\Delta a^{\mathrm{BEC}}$ (m/s$^2$) & 74, 20, 4.8 \\
  Sensitivity $\times 10^{-17}$ & $\Delta G$ (N\,m$^{2}$/kg$^{2}$) & 37, 10, 2.3 \\
  Sensitivity $\times 10^{-31}$ & $\Delta \Lambda$ (m$^{-2}$) & 25, 6.7, 1.6 \\
  \hline
 \end{tabular}
 \caption{Experimental parameters and resulting sensitivities for $^{87}$Rb BEC. Sensitivities to $G$ and $\Lambda$ use $R_0=100$~mm, $\delta_R=1$~mm, and $M=100$~g via Eq.~\eqref{eq:delta-G-L}.}
 \label{table1}
\end{table}
The sensitivities to $G$ and $\Lambda$ can be derived from the acceleration sensitivity $\Delta a^{\mathrm{BEC}}$ using
\begin{equation}
 \Delta G = \Delta a^{\mathrm{BEC}} \frac{R_0^{3}}{2 M \delta_R}, \quad \Delta \Lambda = \Delta a^{\mathrm{BEC}} \frac{3}{\delta_R c^{2}}.
 \label{eq:delta-G-L}
\end{equation}

The relative uncertainty in $G$ is estimated to be $\Delta G / G \sim 10^{-6}$, improving on the current precision of $\sim 10^{-5}$ \cite{Mohr:2024kco}, assuming experimental uncertainties in $\delta_R$, $R_0$, and $M$ are below $10^{-6}$, achievable in vacuum environments \cite{Mohr:2024kco}.
A special discussion regarding the sensitivity to $\Lambda$ is warranted. As shown in Table~\ref{table1}, the expected sensitivity to $\Lambda$ is of the order of $10^{-30}\,\mathrm{m}^{-2}$, representing unprecedented accuracy for Earth-based experiments. Nevertheless, this sensitivity should be regarded as an upper bound, still about 20 orders of magnitude above the actual value measured by the Planck experiment,
\begin{equation}
 \Lambda = (1.09 \pm 0.028) \times 10^{-52} \, \mathrm{m}^{-2} \,,
\end{equation}
as reported in~\cite{Planck:2018vyg}.\footnote{Converted from natural units to SI.}

\begin{figure}[htbp] 
 \centering
 \includegraphics[width=0.48\textwidth]{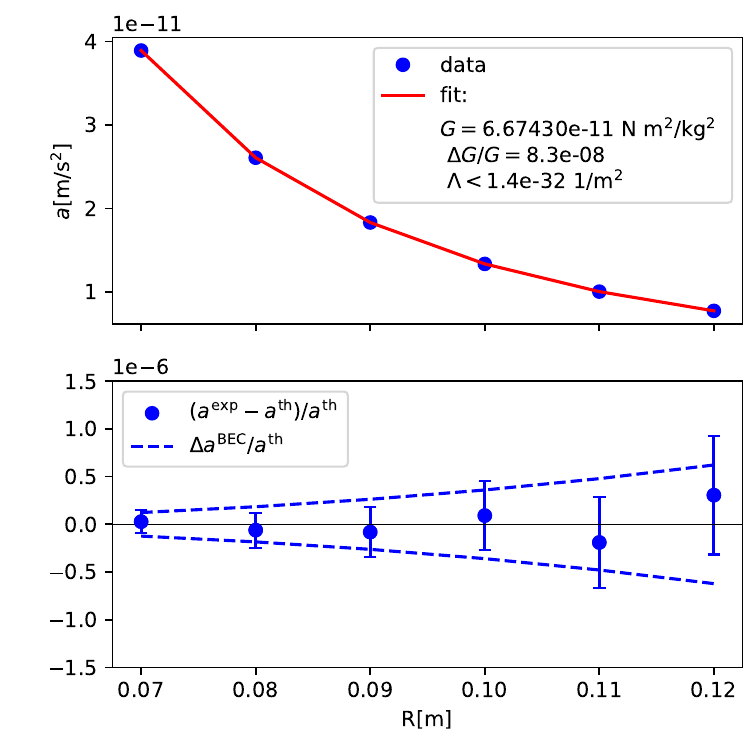}
 \caption{
  \textbf{Top panel:} Simulated measurements of $a^{\mathrm{BEC}}$ (blue circles) as a function of $R_0$, with the fit (red curve) used to determine $G$ and constrain $\Lambda$ assuming sensitivity $\Delta a^{\mathrm{BEC}} = 4.8 \times 10^{-18}$~m/s$^{2}$. 
  \textbf{Bottom panel:} Relative deviation $(a^{\mathrm{exp}} - a^{\mathrm{th}})/a^{\mathrm{th}}$ (blue circles) and expected relative accuracy $\Delta a^{\mathrm{BEC}} / a^{\mathrm{th}}$ (blue dashed line).
 }
 \label{fig:final}
\end{figure}

Figure~\ref{fig:final} displays simulated measurements of the acceleration $a^{\mathrm{BEC}}$ as a function of $R_0$ (the distance between the BEC and the source mass), along with the best-fit curve used to extract $G$ and constrain $\Lambda$. The assumed experimental sensitivity is $\Delta a^{\mathrm{BEC}} = 4.8 \times 10^{-18}$~m/s$^{2}$. The fit utilises the functional dependence of the acceleration on $R_0$ to improve parameter estimation accuracy.

This analysis demonstrates that using a phonon interferometer in a $^{87}$Rb BEC can provide highly precise measurements of Newton's gravitational constant $G$ and place bounds on the cosmological constant $\Lambda$, although current sensitivity to $\Lambda$ remains far from the cosmological value.

\section{Conclusions}
\label{sec:conclusions}

We propose an Earth-based table-top experiment to precisely test the gravitational constant $G$ and the cosmological constant $\Lambda$ using a detector concept that exploits the dynamics of quantum phononic excitations in a trapped Bose--Einstein condensate (BEC), with sensitivity enhanced by approximately 2 orders of magnitude through a tritter operation.

As demonstrated by Gurzadyan's theorem, the fundamental constants we aim to test define the most general functional form of the gravitational potential that satisfies the equivalence between the gravitational forces of a spherical mass and a point mass. The proposed experiment seeks to measure $G$ with an accuracy roughly 2 orders of magnitude better than current measurements and to establish the most stringent Earth-based upper limit on $\Lambda$, thereby pioneering the first laboratory-based probe of the cosmological constant.

Our approach investigates the response of the collective modes of a BEC subjected to an oscillating mass. Using quantum metrology, we derive an expression for the sensitivity in measuring the acceleration amplitude of the time-dependent component of the gravitational potential. This analysis assumes a probe state where the condensed atoms are in a coherent state and two phonon modes are prepared in a two-mode squeezed state.

For the experimental parameters detailed in Table~\ref{table1}, we find that the proposed experiment is sensitive to accelerations on the order of $\simeq 10^{-17}~\mathrm{m/s}^2$, representing an improvement of about 2 orders of magnitude in the current accuracy of $G$ measurements, achieving $\Delta G/G \simeq 10^{-7}$. The experiment would also establish the first Earth-based experimental upper bound on the cosmological constant $\Lambda$ at approximately $10^{-31}~\mathrm{m}^{-2}$. Moreover, this setup enables measurement of the distance-dependent behaviour of each term in the gravitational potential, offering a new level of sensitivity for testing modified gravity theories.

\begin{acknowledgments}
We acknowledge discussions of the results with Roger Penrose.
We thank Hendrik Ulbricht for valuable discussions, including the insight on the level uncertainties for $\delta_{R}$, $R_0$, and $M$ parameters.
We thank Francesco Shankar for the insightful discussions and invaluable contributions, which greatly inspired and influenced the direction of this work.
HF acknowledges that this project was made possible through the support of CONAHCyT. IF thanks an anonymous US philanthropist, John Moussouris, Jussi Westergren, and the Emmy Network for support and research funding. AB and  IF acknowledge support from the Leverhulme Trust project MONDMag (RPG-2022-57). VG thanks AB and IF for hospitality at the University of Southampton. AB is supported in part through the NExT Institute and STFC Consolidated Grant No. ST/L000296/1.
\end{acknowledgments}

\section*{Data availability.}
The data and Python scripts used to generate and analyse the results of this study are publicly available at 
\url{https://github.com/alexanderbelyaev/Probing-Lambda-Gravity-python}. 
Additional information is available from the corresponding author upon reasonable request.

\appendix
\setcounter{section}{0}

\label{supp:EM-aBEC}

\section{Gravitational Potential and Acceleration Induced by an Oscillating Mass in the BEC}
\label{supp:grav-pot}
The gravitational potential generated by the oscillating mass in BEC at point $x$
as follows from Eqs.~3 and 4 is given 
by
\begin{equation}
 \phi(x,t)=-\frac{M G}{r(x,t)} - \frac{\Lambda r(x,t)^2c^2}{6}.
\end{equation}
The expansion of  $\phi(x,t)$ around $R_0$ 
up to the second order in  $\Delta_{x,t}$ leads to the following expression for the $\phi(x,t)$:
\begin{equation}
 \phi(x,t)
 = \phi_0^G(1-\Delta_{x,t}+\Delta_{x,t}^2 + ...) +
 \phi_0^\Lambda(1+ 2\Delta_{x,t}+\Delta_{x,t}^2)
 \label{eq:phi}
\end{equation}
where $\phi_0^G=-\frac{M G}{R_0} $ and 
$\phi_0^\Lambda = -\frac{\Lambda R_0^2c^2}{6}$.
The part of $\phi(x,t)$ from Eq.~\ref{eq:phi} to which the BEC experiment has essential sensitivity (as we discuss below) is proportional to $[x \delta_R \sin(\Omega t)]$, originating from the $\Delta_{x,t}^2$ and higher-order $\mathcal{O}(\Delta_{x,t}^2)$ terms, and is given by
\begin{equation}
 \phi^\text{BEC}_\Omega(x,t)=
 \frac{2 x \delta_R  }{R_0^2}\left(\phi_0^G(1+...) + \phi_0^\Lambda\right)\sin(\Omega t)
 \equiv a^\text{BEC}_\Omega(t) x,
 \label{eq:phi-bec}
\end{equation}
where `...' denotes 
$\mathcal{O}\left(\frac{\delta R^2}{R_0^2}\right)$ terms, and
\begin{equation}
 a^\text{BEC}_\Omega(t)=
 \frac{2\delta_R}{R_0^2} \left(\phi_0^G (1+ ...) + \phi_0^\Lambda\right)\sin(\Omega t),
 \label{eq:a-bec}
\end{equation}
which is the time-dependent part of the acceleration exerted by the oscillating sphere on the BEC.
The amplitude of this acceleration, $a^\text{BEC}$, is the key observable in our study, which follows from the two equations above and is given by
\begin{equation}
 a^\text{BEC} \simeq 
 \frac{2\delta_R}{R_0^2}\left(\frac{MG}{ R_0}  + \frac{\Lambda R_0^2 c^2}{6}\right),
\end{equation}
where we have omitted the $\mathcal{O}(\frac{\delta R^2}{R_0^2})$ corrections to the first term with constant $G$.

\section{Quantum Field Treatment of BEC Phonons} 
\label{supp:QFBEC}

The standard description of a Bose gas with two-atom interactions \cite{Pitaevskii2003BoseE}, taking into account the gravitational potential of the oscillating sphere, is given by

\begin{align}
 \hat{H} &= \int_{\mathcal{V}} \hat{\Psi}^{\dagger} \left( - \frac{\hbar^{2}}{2m} \nabla^{2} +  V_{\textrm{tr}} -  m \phi(x,t)  + \frac{g}{2} \hat{\Psi}^{\dagger }  \hat{\Psi} \right) \hat{\Psi} \, d^{3}x ,
 \label{eq:h-hat}
\end{align}

\noindent where $V_{\textrm{tr}}$ is the trapping potential, $g = 4 \pi \hbar^{2} a_{sl} /m$ is the two-atom coupling constant, $m$ the mass of the atoms in the BEC, $a_{sl}$ the atomic scattering length and $\mathcal{V}$ is the confinement volume over which  Eq.~\ref{eq:h-hat} is integrated. To produce a one-dimensional uniform density BEC in the $x$-direction, we set $V_{\textrm{tr}}=0$ inside the trap and impose Neumann boundary conditions at the potential walls. 
The stationary part of the gravitational potential $\phi(x,t)$, as well as any $x$-independent terms, do not contribute to the system's dynamics. The only term contributing to the gravitational interaction is given by  $\phi^\rm{BEC}_\Omega(x,t)=a^\rm{BEC}_\Omega(t) x$, with $a^\rm{BEC}_\Omega(t)$ given by Eq.~\ref{eq:a-bec} \cite{Ratzel2018Dynam}. The field operator can be expanded as
\begin{equation}
 \label{IV.02}
 \hat{\Psi} (\textbf{r}, t) = (\hat{\Psi}_{0} (\textbf{r}) + \hat{\vartheta} (\textbf{r}, t) ) e^{-i\mu t/ \hbar - i \int_{0}^{t} \delta \mu (t') dt' / \hbar} ,
\end{equation}

\noindent where $\hat{\Psi}_{0} (\textbf{r})$ is the solution of the stationary Gross-Pitaevskii equation, $\hat{\vartheta} (\textbf{r}, t)$ is a small perturbation, $\mu$ is the chemical potential and {$\delta\mu = \int \phi^\text{BEC}_\Omega(t) d^{3}x$} is the time-dependent energy shift of the ground state.

Bose-Einstein condensation is achieved assuming that the temperature $T$ of the Bose gas is much smaller than the condensation's critical temperature so that the ground state becomes macroscopically occupied. Making the \textit{Bogoliubov approximation}, we can replace the field operator with a classical mean-field function $\hat{\Psi}_{0} (\textbf{r}) = \hat{a}_{0} \, \psi_{0}(\textbf{r}) \rightarrow \sqrt{N_{0}} \psi_{0}(\textbf{r})$, where $N_{0}$ corresponds to the number of atoms in the ground state of the BEC. The perturbations are $\hat{\vartheta} (\textbf{r}, t) = \sum_{n \neq 0} \hat{a}_{n} (t) \, \psi_{n}(\textbf{r})$, where $\hat{a}_n^{\dagger}$ and $\hat{a}_n$ are the creation and annihilation atom operators, satisfying the commutation relation $[\hat{a}_{n},\hat{a}_{l}^{\dagger}] = \delta_{nl}$. To help solve the equations of motion, it is convenient to apply the \textit{Bogoliubov transformation}
\begin{equation}
 \label{IV.05}
 \hat{\vartheta}(\textbf{r}, t) = \sum_{n} \left( u_{n} (\textbf{r}) \hat{b}_{n} e^{-i \omega_{n} t} + v_{n} (\textbf{r}) \hat{b}_{n}^{\dagger} e^{i \omega_{n} t} \right) ,
\end{equation}

\noindent where $\hat{b}_{n}^{\dagger}$ and $\hat{b}_{n}$ are the Bogoliubov mode creation and annihilation operators obeying $[\hat{b}_{n}, \hat{b}_{l}^{\dagger}] = \delta_{nl}$ and $\omega_{n}$ is the corresponding mode frequency. The mode functions $u_{n}(x)$, $v_{n}(x)$ follow the stationary Bogoliubov-de-Gennes equations and satisfy the orthogonality relation  $\int (u_{n}^{*}u_{l} - v_{n}^{*}v_{l}) \, d^{3}x = \delta_{nl}$.

The energy spectrum is given by the dispersion relation
\begin{equation}
 (\hbar \omega_{n})^{2} = (c_{s} \hbar k_{n})^{2} + (\hbar^{2} k_{n}^{2} / 2m)^{2} ,
\end{equation}
where 
\begin{equation}
 \label{eq:cs}
 c_{s}: = \sqrt{g n_{0} / m}
 =\sqrt{4\pi a_{sl} n_{0}}\hbar/m
\end{equation}
is the speed of sound, $n_{0}$ is the BEC's density and $k_{n} = n\pi/L$ the mode number with $n\in\mathbb{Z}^{+}$. In the \textit{low-energy limit} ($\hbar \omega_{n} \ll m c_{s}^{2}$), the dispersion relation is $\omega_{n} = c_{s} k_{n}$. The Bogoliubov modes in this limit correspond to phonons. 
In the interaction picture, we can rewrite the Hamiltonian as the sum of a diagonal part, $\hat{H}^{(0)} = \sum_{n}: \hbar \omega_{n} \hat{b}_{n}^{\dagger} \hat{b}_{n}: $ (where $:\,:$ denotes normal ordering), plus an interaction term $\hat{H}^{(I)}$, that will be specified in what follows. 
We will consider resonant effects between the phononic modes and the oscillation frequency of the mass. In particular, we consider that the mode numbers satisfy the condition $n_{\Omega}:=n + l = L\Omega/(\pi c_{s})$, with $n_\Omega$ an odd integer, which corresponds to the resonance condition $\Omega \approx \omega_{n} + \omega_{l}$. In this case, the main contribution from the interaction Hamiltonian up to second order in the phonon operators under the rotating wave approximation \cite{Ratzel2018Dynam}
\begin{align} 
 \label{14}
 \hat{H}^{(I)} =  - \sum_{l < n_\Omega} i(-1)^{n_\Omega} |M_{nl}| a^\rm{BEC} \left( \hat{b}_{n}^{\dagger} \hat{b}_{l}^{\dagger} - \hat{b}_{n} \hat{b}_{l} \right) ,
\end{align}

\noindent for $n \neq l$, where the transition amplitude is given by

\begin{align}
 |M_{nl}| \approx \frac{ m L^{2} (n^{2}+l^{2}) (1-(-1)^{n_\Omega})}{2 \sqrt{2nl} (n^{2} - l^{2})^{2} \pi^{3} \zeta},
\end{align}

\noindent where $\zeta = \hbar / (\sqrt{2} m c_s)$ is the healing length. 

The time evolution of the phonon modes is given by the operator $\hat{U}(t) = \exp{(i\hat{H}^{(I)}t/\hbar)}$, which explicitly reads
\begin{align}
 \label{eq:UT}
 \hat{U} (t) = \exp\left[ - \sum_{l<n_{\Omega}} (-1)^{n_{\Omega}} a^\rm{BEC} |M_{nl}| \left( \hat{b}_{n}^{\dagger} \hat{b}_{l}^{\dagger} - \hat{b}_{n} \hat{b}_{l} \right)t/\hbar \right]  .
\end{align}

\noindent This unitary operator corresponds to a two-mode squeezing transformation, parameterized by $a^\rm{BEC}$. 

\section{Inverse Tritter Operation}
\label{sec:inverse-tritter}

The tritter Hamiltonian, inspired by beam-splitting-like couplings applied to the BEC's ground state and phonon modes, is given by
\begin{equation*}
\hat{H}_{\text{tr}} = \hbar \theta \sqrt{2} \left[ e^{i\vartheta} \hat{a}_0^\dagger (\hat{b}_n + \hat{b}_l) + e^{-i\vartheta} \hat{a}_0 (\hat{b}_n^\dagger + \hat{b}_l^\dagger) \right],
\end{equation*}
\noindent which can be realized through suitable modulation of the trapping potential or through Bragg diffraction \cite{PhysRevLett.82.4569, Szigeti_2012}. While experimental demonstrations of beam-splitting operations are available, the inverse operation is less commonly reported. The corresponding unitary operator $\hat{U}_{\text{tr}} = \exp(-i \hat{H}_{\text{tr}} t / \hbar)$ can be inverted by changing the sign of the phase $\vartheta \rightarrow \vartheta + \pi$, thus flipping the sign of the Hamiltonian. This is analogous to time-reversing a beam splitter in linear optics. In cold-atom setups, such coupling between the condensate and phonon modes can be implemented via Bragg diffraction using two counter-propagating laser beams tuned to excite phonons from the ground state or return the excitations to the ground state. Controlling the phase of these beams enables reversal of the interaction Hamiltonian, thereby implementing $\hat{U}_{\text{tr}}^{-1}$.

These mechanisms, while technically challenging, are within reach of current experimental capabilities and form a promising direction in quantum sensing and information protocols using phononic degrees of freedom.

\bibliography{refs}
\bibliographystyle{unsrtnat}

\end{document}